\documentclass[aps,preprint,showpacs]{revtex4}

\usepackage{graphicx}

\begin{document}

\title{Ferromagnetic instabilities in neutron matter at finite
       temperature with the Skyrme interaction}

\author{Arnau Rios and Artur Polls }

\affiliation{Departament d'Estructura i Constituents de la Mat\`eria,
         Universitat de Barcelona, E-08028 Barcelona, Spain}

\author{Isaac Vida\~na }

\affiliation{ Gesellschaft f\"ur Schwerionenforschung (GSI), Planckstrasse 1, D-64291 Darmstadt,
Germany }

\begin{abstract}
The properties of spin polarized neutron matter are studied both at zero and finite temperature using
Skyrme-type interactions. It is shown that the critical density at which ferromagnetism takes place decreases
with temperature. This unexpected behaviour is associated to an anomalous behaviour of the entropy which becomes
larger for the polarized phase than for the unpolarized one above a certain critical density. 
This fact is a consequence of the dependence of the entropy on the effective mass of the neutrons
with different third spin component. A new constraint on the parameters of the effective Skyrme 
force is derived if this behaviour is to be avoided.
\end{abstract}
\vspace{0.5cm}

\maketitle


\section{Introduction}

Since the pioneering work of Vautherin and Brink \cite{brink1972} 
there has been an intensive use of Skyrme effective nucleon-nucleon
(NN) interactions to study properties of finite nuclei and 
nuclear matter, the latter one mainly in conditions of astrophysical interest. 
This type of phenomenological NN interaction
is thought to  be used within a Hartree-Fock scheme (HF). Its zero-range
 character leads to simple analytical expressions
for basic properties of symmetric nuclear matter such as the binding 
energy ($a_v$), the saturation density ($\rho_0$), the incompressibility modulus 
($K_{\infty}$) or the symmetry energy ($a_s$).
Actually, the experimental values of these quantities together with
the binding energy of some doubly magic nuclei have been traditionally
 used to fit the parameters entering the general expression of a Skyrme force. 
The main advantage of these forces comes from their 
analytical character, which makes them very useful to get a physical insight
 into problems where the fully microscopic calculations are either very time 
consuming or not yet possible to implement. 
Once a parametrization is determined,  
 one can, for instance, study nuclei and nuclear matter under conditions 
far from those used to fix the force in quite a simple way. In doing so, however, 
one must always keep in mind the particular limitations of the Skyrme 
parametrization which is being used.  

By construction, most of the Skyrme forces used in the literature are well 
behaved around the saturation density of nuclear matter and for moderate 
isospin asymmetries. However, not all the Skyrme parameters are completely 
well determined through the fits of given sets of data and only certain combinations
related to the basic properties mentioned above are really empirically determined \cite{margue2002}.
This leads to 
a scenario where, for instance, different Skyrme forces produce similar 
equations of state for symmetric nuclear matter but very different 
results for neutron matter. 
This is easily understood if one considers that neutron matter
or, equivalently, the systems with large isospin asymmetries, are not part of 
the common input data that determine the parameters of the interaction.  
Obviously, this feature should be corrected if Skyrme-type forces are to 
be used in conditions of large neutron to proton ratios such as nuclear matter
 inside neutron stars or nuclei near the drip line. Recently, several sets of 
Skyrme forces have been constructed
taking into account different data coming from highly isospin asymmetric systems. 
The most well-known among these are probably the parametrizations of the Lyon 
group (SLy interactions) \cite{sly1,sly2}, which also took into account variational 
results for neutron matter obtained with a realistic interaction \cite{fabro88}
 to fix the parameters of the force. Another important piece of input for
 neutron-rich systems are the isotope shifts of medium and heavy nuclei. 
In that case, 
modifications of the spin orbit term of the force have been taken into account to correctly 
reproduce the data. With these prescriptions, the SkI parametrizations were 
created almost a decade ago \cite{ski}. 
Recently, an extensive and systematic study has tested
 the capabilites of almost ninety existing Skyrme parametrizations to provide 
good neutron-star properties \cite{stone2003}. It was found that only twenty seven of these
 forces passed the restrictive tests imposed, the key property being the  
behaviour (increasing) of the symmetry energy $a_s$ with density. 

Another situation of astrophysical interest refers to the possibility of the
spontaneous appearance of spin polarized states in nuclear matter. This type
of instability, {\it i.e.,} a ferromagnetic transition at high densities, has been 
studied since long ago using different theoretical methods 
\cite{br69,ri69,cc69,cl69,si69,os70,pe70,pa72,ba73,ha75,ja82,ma91,ku89,be95}. The results are,
nevertheless, still contradictory. On the one hand, 
Skyrme interactions predict several types of instabilities at increasing density 
\cite{margue2002}. In particular, currently used Skyrme forces show
a ferromagnetic transition for neutron matter at densities in the range (1.1-3.5)$\rho_0$ 
\cite{nava84,kuts94}.
However, it has recently been shown that by including a
small fraction of protons into the system, the onset density of ferromagnetism
can be substantially reduced \cite{isayev2004a}.   
On the other hand, recent Monte Carlo simulations 
\cite{fan2001} and also Brueckner-Hartree-Fock calculations \cite{vida2002,vida2002b}
 using modern two- and three-body realistic interactions exclude such an
instability, at least at densities up to five or six times $\rho_0$. 
This transition could 
have important consequences for the evolution of a protoneutron star, in
 particular for the spin correlations in the medium which do strongly affect
the neutrino cross sections and the neutrino mean free paths inside the star 
\cite{nava1999}.
Therefore, drastically different scenarios for the evolution of protoneutron 
stars are to be considered if the existence of such a ferromagnetic transition
 is confirmed.

Most of the studies of the ferromagnetic instability have been conducted at 
zero temperature. However, the description of protoneutron stars requires a 
study at finite temperature. Thus, since the general conditions for ferromagnetism are 
well established in the case of the Skyrme forces at zero temperature and even though their 
predictions are quite different than those of microscopic calculations,  it is
still interesting to study how the situation is modified at finite temperature.
Intuitively, due to the thermal disorder, one would expect the onset density for 
ferromagnetism to increase with temperature. But, as we will see later on,
the behaviour turns out to be the opposite one and what actually happens is 
that the ferromagnetic transition takes place at smaller densities. In this 
work, we associate this fact to an anomalous behaviour of the entropy as a 
function of the spin polarization.
However, it is worth noticing here that there is no thermodynamical inconsistency in this dependence. If
eventually, microscopic calculations with realistic interactions would not confirm this 
trend, we find out a new constraint on the parameters of the effective Skyrme force 
in order to avoid such a behaviour.

In the next section we summarize the expressions of the energy and the 
single-particle energies at zero temperature and  find
the free energy and entropy at finite temperature for neutron matter as a 
function of the spin polarization. In section III, we present some 
results involving the thermodynamical potentials of the system. Section IV 
is devoted to the study of the entropy, its relation to the effective
mass and the possible constraints which could be imposed to the parameters in
order to avoid such an anomalous behaviour in both the classical and the low temperature
limit. Finally, the main conclusions are summarized in section V.

\section{Polarized neutron matter}

Most of the Skyrme interactions used in the literature have the following general form:
\begin{eqnarray}
V({\bf r_1}, {\bf r_2})& =& t_0 (1 + x_0 P^{\sigma}) \delta({\bf r}) +
\frac {1}{6} t_3 (1 + x_3 P^{\sigma}) \left [\rho({\bf R}) \right ]^{\alpha} \delta({\bf r}) \nonumber
 \\
 & & + \frac {1}{2} t_1 ( 1 + x_1 P^{\sigma}) ( {\bf k'}^2 \delta({\bf r}) +
\delta({\bf r}) {\bf k}^2) +
t_2 ( 1 + x_2 P^{\sigma}) {\bf k}'\cdot \delta({\bf r}) {\bf k} \nonumber \\
 &  & + i  W_0 ({\bf \sigma}_1 + {\bf \sigma}_2) \cdot \left [ {\bf k}' \times \delta({\bf r})
{\bf k} \right ]
\label{eq:skyrme}
\end{eqnarray}
with ${\bf r}= {\bf r}_1 -{\bf r}_2$, ${\bf R}= ({\bf r}_1 + {\bf r}_2)/2$, 
${\bf k}= ({\bf \nabla}_1 -{\bf \nabla}_2)/2 i $ the relative momentum
acting on the right and ${\bf k}'$ its conjugate acting on the left. 
$P^{\sigma}= (1 + \vec{\sigma}_1 \cdot \vec{\sigma}_2)/2$ is the spin exchange
operator. The last term, proportional to $W_0$, corresponds to the zero-range spin-orbit term,
 which does not contribute in homogenous systems and thus will be 
ignored for the rest of the paper.

Let us consider a homogeneous system of neutrons characterized by a total
density $\rho$, which is the sum of the  spin-up ($\rho_{\uparrow}$) and  
spin-down ($\rho_{\downarrow}$) densities. At zero temperature, the total
energy will be a function of $\rho_{\uparrow}$ and $\rho_{\downarrow}$ or, 
alternatively, of the total density $\rho$ and the spin polarization parameter
 $\Delta$, defined as $\Delta = (\rho_{\uparrow}-\rho_{\downarrow})/\rho$.

The total energy of this system in the Hartree-Fock approximation is given by 
the sum of the kinetic energy associated to the Fermi gas of polarized 
neutron matter and the expectation value of the Skyrme interaction between the
 wave function describing two free Fermi seas corresponding to neutrons with 
two different spin 
orientations.  These two Fermi seas have all the single-particle states 
characterized by a good linear momentum, occupied up to the Fermi levels
$k_{F\uparrow}$ and $k_{F\downarrow}$ defined through
$k_{F\uparrow (\downarrow)} = ( 6 \pi^2 \rho_{\uparrow (\downarrow)} )^{1/3}$.

The total energy per particle for zero temperature has the following expression:
\begin{eqnarray}
e\left(\rho_{\uparrow},\rho_{\downarrow} \right) &=& \frac{\hbar^2}{2 m} \frac{1}{\rho} \left[ \tau_{\uparrow}+\tau_{\downarrow} \right]
 + \frac{1}{4\rho} \left[2t_2(1+x_2) \right] \left[
\tau_{\uparrow}\rho_{\uparrow} + \tau_{\downarrow}\rho_{\downarrow} \right]
\nonumber \\ 
 && + \frac{1}{4\rho} \left[ t_1(1-x_1)+t_2(1+x_2) \right] \left[
\tau_{\uparrow}\rho_{\downarrow} + \tau_{\downarrow}\rho_{\uparrow} \right]
\nonumber \\ 
&& + \frac{1}{\rho} \left[ t_0(1-x_0)+ \frac{1}{6}t_3(1-x_3) 
\rho^{\alpha} \right] \rho_{\uparrow}\rho_{\downarrow} \; .
\label{eq:ener} 
\end{eqnarray}
To simplify the notation, we will use from now on the symbol $\sigma$ to indicate the third spin component. 
The functions $\tau_{\sigma}$ are related to the average 
kinetic energy of the Fermi model of polarized neutron matter:
\begin{equation}
\tau_{\sigma} = \frac {3}{5} (6  \pi^2 \rho_{\sigma})^{2/3} 
\rho_{\sigma} = \frac {3}{10} \left( 3 \pi^2 \rho \right)^{2/3}
\rho  (1 \pm \Delta)^{5/3} \; ,
\end{equation}
where the plus (minus) sign corresponds to the up (down) spin projection. From the energy per particle  one can easily derive the chemical potentials
 (up and down) and the pressure. Another important quantity for our analysis 
is the single-particle energy:
\begin{equation}
\epsilon_{\sigma}(k) = \frac {\hbar^2 k^2}{2m} + U_{\sigma}
(k,\rho_{\uparrow},\rho_{\downarrow}) \; .
\label{eq:single}
\end{equation}
The single-particle potential 
$U_{\sigma}(k,\rho_{\uparrow},\rho_{\downarrow})$
takes into account the interaction of a particle with momentum ${\bf k}$ and 
spin projection $\sigma$ with all the rest. It has a quadratic 
dependence on the momentum which is usually incorporated in the single-particle spectrum as a momentum independent effective mass:
\begin{equation}
\epsilon_{\sigma}(k) = \frac {\hbar^2 k^2}{2m_{\sigma}^*} + \bar U_{\sigma}(\rho_{\uparrow},\rho_{\downarrow}) \; ,
\label{eq:single1}
\end{equation}
where the effective mass is given by
\begin{equation}
\frac{m^*_{\sigma}}{m}=\frac{1}{ 1 + \frac{2m}{\hbar^2}
 a_{\sigma} \left(
\rho_{\uparrow},\rho_{\downarrow} \right) } \; ,
\label{eq:effec}
\end{equation}
with
\begin{eqnarray}
a_{\sigma} \left(\rho_{\uparrow}, \rho_{\downarrow} \right) 
&=& \frac{1}{4} \Big[
   \left[2t_2(1+x_2)\right] \rho_{\sigma} +
   \left[ t_1(1-x_1)+t_2(1+x_2) \right]\rho_{-\sigma} \Big] \; .
\end{eqnarray}
The momentum independent part $\bar U_{\sigma}$ of the 
single-particle potential $U_{\sigma}$ is then given by:
\begin{eqnarray}
\bar{U}_{\sigma}\left(\rho_{\uparrow},\rho_{\downarrow}\right) &=&
\frac{1}{4} \Big[ \left[2t_2(1+x_2)\right] \tau_{\sigma} +
   \left[ t_1(1-x_1)+t_2(1+x_2) \right]\tau_{-\sigma}\Big] \nonumber \\ 
&&+ \left[ t_0(1-x_0)+\frac{1}{6}t_3(1-x_3) \rho^{\alpha} \right]
 \rho_{-\sigma}
     + \frac{1}{6} \alpha t_3 \rho^{\alpha-1} \rho_{\sigma} \rho_{-\sigma} \; .
\label{eq:uinde}
\end{eqnarray}

It is worth mentioning that the expression for $\bar U_{\sigma}$
 contains also the rearrangement term $U_R(\rho_{\uparrow},\rho_{\downarrow})$:

\begin{equation}
U_R(\rho_{\uparrow},\rho_{\downarrow})= \frac {1}{6} \alpha t_3 \rho^{\alpha -1} \rho_{\sigma}
\rho_{-\sigma} ,
\label{eq:reare}
\end{equation}
which takes into account the effect of the density dependence of the effective 
interaction on the single-particle potential. 
Using this prescription for the single-particle potential, one can check 
that the chemical potential calculated from the energy per particle: 
\begin{equation}
\mu_{\sigma}(\rho_{\uparrow},\rho_{\downarrow}) = e(\rho_{\uparrow},\rho_{\downarrow})+ \rho \left (\frac {\partial
e(\rho_{\uparrow},\rho_{\downarrow}) }{\partial \rho_{\sigma}} \right )_{
\rho_{-\sigma}} \; ,
\end{equation}
does exactly coincide with the single-particle energies at the respective Fermi
 surfaces of each species, 
$ \mu_{\sigma}=\epsilon(k_{F \sigma })$.

The extension of these expressions to finite temperature is rather 
straightforward. The expression for the internal energy and for the 
single-particle energy are exactly the same as in the zero temperature case, 
the only change coming from $\tau_{\sigma}$,
 which at finite temperature is given in terms of the so-called Fermi 
integrals \cite{pathria}:
\begin{equation}
\tau_{\sigma}=\frac{g}{(2\pi)^2} 
\left( \frac{2m_{\sigma}^*}{\hbar^2} T\right)^{5/2}
J_{3/2}(\eta_{\sigma}) \; ,
\end{equation}
where $g$ is the spin degeneracy factor of the system (in our case $g=1$ for each 
spin component) and where:
\begin{equation}
J_{\nu}(\eta)=\int_0^{\infty} dx \frac{x^{\nu}}{1+e^{x-\eta}} \; .
\end{equation}
The parameter  
$\eta_{\sigma}$ should be calculated by inverting the equation:
\begin{equation}
\rho_{\sigma}=\frac{g}{(2\pi)^2} 
\left( \frac{2m_{\sigma}^*} {\hbar^2} T\right)^{3/2}
J_{1/2}(\eta_{\sigma}) \; ,
\label{norm}
\end{equation}
which simply states, in terms of $\eta_{\sigma}$, that the
integration of the Fermi momentum distribution of each spin component:
\begin{equation}
n_{\sigma}(k) = \left ( 1 + 
e^{\frac {\epsilon_{\sigma} (k) - \mu_{\sigma}}{T}} \right )^{-1}
\end{equation}
should coincide with the density of the component. 

Once $\eta_{\sigma}$ has been obtained, one can also calculate
 the chemical potentials at finite temperature:
\begin{equation}
\mu_{\sigma} (\rho_{\uparrow},\rho_{\downarrow},T) = \eta_{\sigma} T + \bar U_{\sigma} (\rho_{\uparrow},\rho_{\downarrow},T)
\end{equation}
and the entropy per particle:
\begin{equation}
s(\rho_{\uparrow},\rho_{\downarrow},T) =
\frac {\rho_{\uparrow}}{\rho} s_{\uparrow} (\rho_{\uparrow},\rho_{\downarrow},T) +
\frac {\rho_{\downarrow}}{\rho} s_{\downarrow}(\rho_{\downarrow},\rho_{\uparrow},T) \;
\end{equation}
where the entropies of each component are given in the Hartree-Fock approximation by:
\begin{eqnarray}
s_{\sigma}(\rho_{\uparrow},\rho_{\downarrow},T)&=& 
- \frac{1}{\rho_{\sigma}}\int \frac{d^3k}{(2\pi)^3} \left[ n_{\sigma}(k) \ln n_{\sigma}(k) +
(1- n_{\sigma}(k)) \ln(1- n_{\sigma}(k)) \right] \nonumber \\
&=& \frac{5}{3} \frac{1}{\rho_{\sigma}}\frac{1}{4\pi^2} \left( \frac{2m^*_{\sigma}}{\hbar^2} T \right)^{3/2} J_{3/2}(\eta_{\sigma}) - \eta_{\sigma} \; .
\end{eqnarray}

Finally, for a fixed density and temperature, the suitable thermodynamical potential
is the free energy. We can get if from the previous expressions of the energy and the
 entropy:
\begin{equation}
f(\rho_{\uparrow},\rho_{\downarrow},T ) = e(\rho_{\uparrow},\rho_{\downarrow},T)- T
s(\rho_{\uparrow},\rho_{\downarrow}, T) \; .
\end{equation}
From the free energy per particle, we can get the rest of the macroscopic properties of the 
system as, for instance, the pressure (and thus the EoS). In our case, we are 
particularly interested in the inverse magnetic susceptibility $\chi^{-1}$, which can be 
obtained from a second derivative of the free energy with respect to the spin polarization:
\begin{equation}
\frac{1}{\chi}=\frac{1}{\mu^2 \rho} 
\left( \frac{\partial^2 f}{\partial \Delta^2} \right)_{\Delta=0} \; ,
\label{eq:mag1}
\end{equation}
where $\mu$ is the magnetic moment of the neutron. 

Notice that in our approach the effective mass $m^*$ does not depend on the 
temperature and that the chemical potential obtained from the 
normalization condition of the density of each component (\ref{norm}), when the
 rearrangement is taken into account, coincides with the chemical potential 
derived from the free energy through its derivative with respect to density.

\section{Energetics of polarized neutron matter}

For the discussion of our results we have chosen the SLy4 \cite{sly2} and the SkI3 \cite{ski} Skyrme forces, both
of which passed successfully the careful 
tests of Ref.\ \cite{stone2003}.
In Table \ref{table1}, we report the values at saturation density of the binding energy, symmetry energy,
 incompressibility modulus and maximum mass of the neutron star which can be obtained using
the EoS of these two Skyrme forces when the $\beta$- stability conditions in the presence of electrons and muons are imposed. 

\begin{figure}[thb]
   \includegraphics[height=10cm,width=12cm]{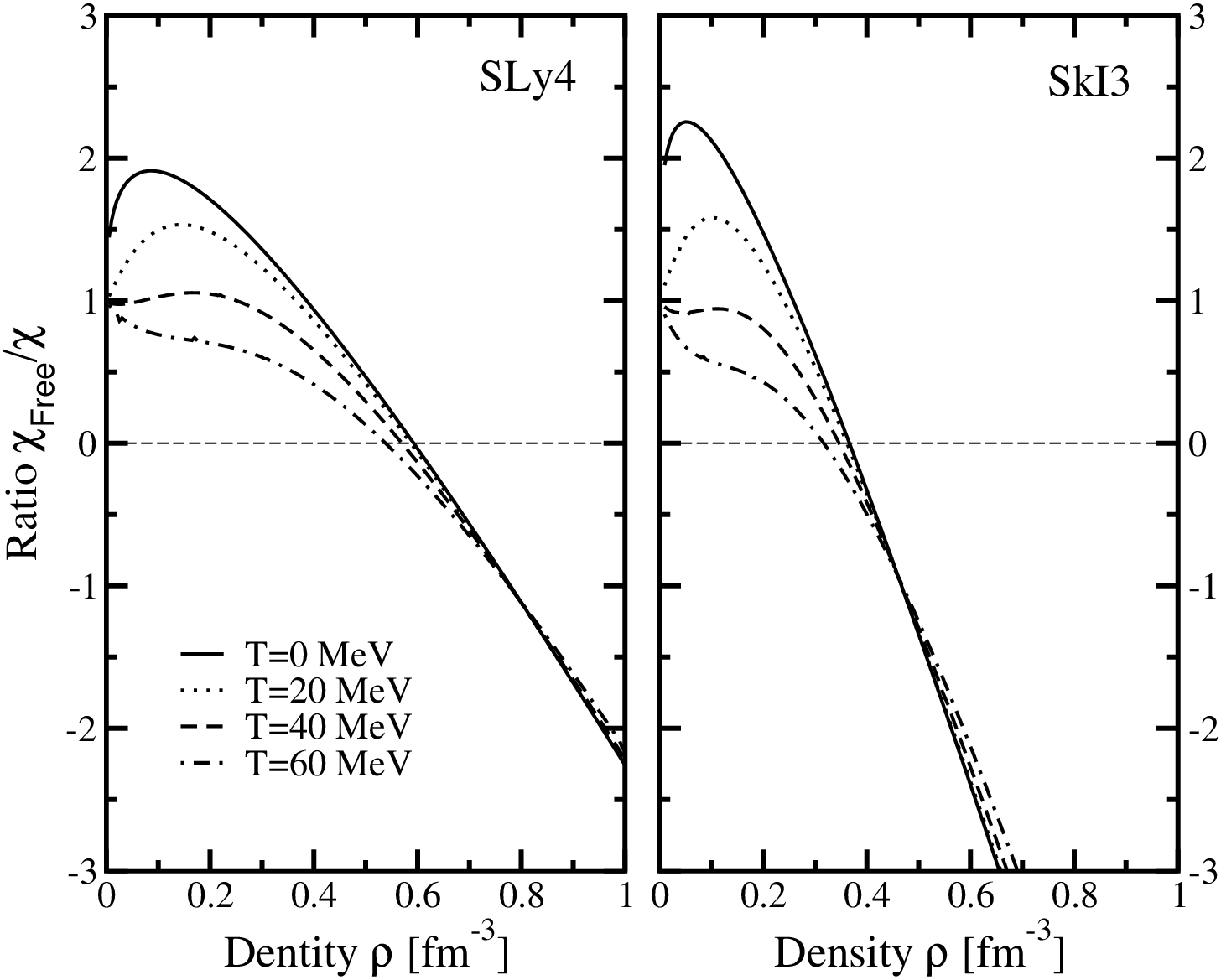}
   \vspace{0.75cm}
   \caption{Ratio between the magnetic susceptibility of the free Fermi gas and  
			the corresponding magnetic susceptibility of interacting neutron matter 
                        as a function of density for several temperatures.}
   \label{fig:fig1}
\end{figure}

Since we are interested in analyzing the behaviour of these forces with respect
to the polarization of neutron matter, in Fig.\ (\ref{fig:fig1}) we report the 
ratio between the inverse magnetic susceptibility of interacting neutron matter
and that of the corresponding free Fermi gas of neutrons as a function of the 
density for different temperatures. At zero temperature,
as it is well known, both forces present a magnetic instability, {\it i.e.,} the 
previous ratio becomes zero at the critical densities ($\rho_c=0.60 \mbox{ fm}^{-3}$ 
for SLy4, $\rho_c=0.37 \mbox{ fm}^{-3}$ for SkI3), but, contrary to what can
be intuitively expected, the onset density for the magnetic instability 
decreases with temperature for both forces. It is precisely this anomalous behaviour
that we want to explore here. Since we are mainly interested on the study of thermal 
effects, we have explored unrealistically high temperatures (much higher than
those needed in the evolution of proto-neutron stars) in order to magnify them.

\begin{figure}[thb]
   \includegraphics[height=10cm,width=12cm]{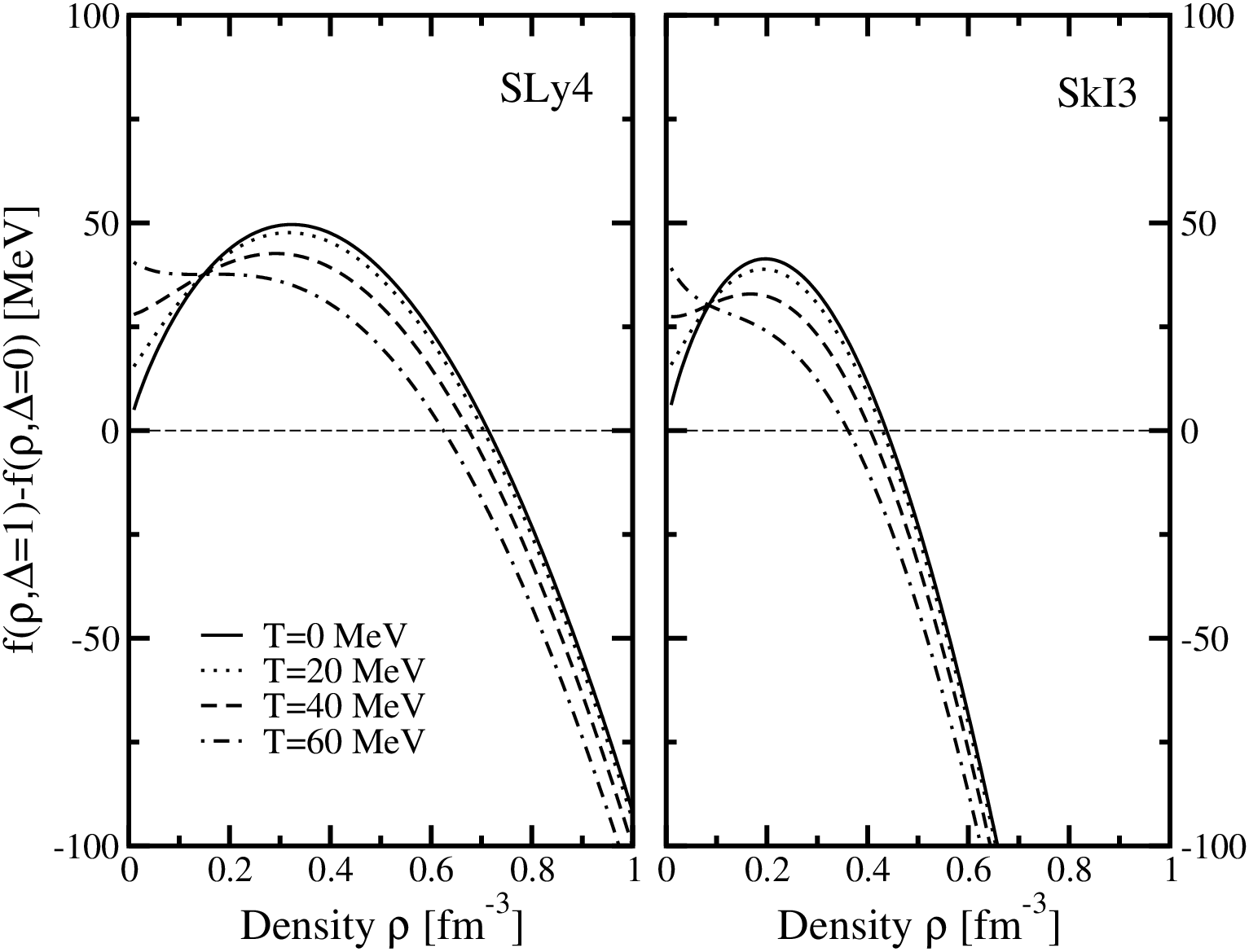}
   \vspace{0.75cm}
   \caption{ Difference between the free energy per particle of fully polarized neutron matter and unpolarized neutron matter as a function of density for several temperatures.}
   \label{fig:fig2}
\end{figure}

A complementary information which is rather helpful in this analysis is the difference
between the free energy of totally polarized and unpolarized neutron matter. This difference 
as a function of density is reported in Fig.\ (\ref{fig:fig2}). Once this difference has become negative,
the totally polarized system will have a lower free energy and therefore the system will prefer
a polarized phase in front of the non-polarized one.
Notice that the density at which this difference becomes negative
does not coincide with the onset of the magnetic instability. The critical densities 
defined by this criterium at zero temperature are $\rho_c^F = 0.71 \mbox{ fm}^{-3}$ 
and $\rho_c^F= 0.44 \mbox{ fm}^{-3}$ respectively, both of them larger than the corresponding $\rho_c$'s. 
The critical density $\rho_c$ signals the density at which the unpolarized phase becomes 
unstable around $\Delta=0$, which, however, does not imply that the system prefers
the fully polarized phase, {\it i.e.,} the minimum of the free energy is not necessarily located at $\Delta=1$. 
Even more, beyond $\rho_c^F$, as we will check later on, one can not guarantee that the minimum
of the free energy is located in the fully polarized system.

In this discussion, it is worth noticing that the energy per particle for fully 
polarized neutron matter with Skyrme interactions reduces to: 
\begin{equation}
e(\rho,T,\Delta=1) = \left [\frac {\hbar^2}{2 m} \frac {1}{\rho}+
\frac {1}{2} t_2 (1+x_2) \right ] \tau_{\uparrow}= \frac {\hbar^2}{2 m^*}
\frac {1}{\rho} \tau_{\uparrow} \; ,
\label{eq:neutpola}
\end{equation}
where the effective mass $m^*$ depends only on the parameters 
$t_2$ and $x_2$:
\begin{equation}
\frac {m^*(\rho,\Delta=1) }{m} = \frac {1} {1 + \frac {2m}{\hbar^2} \frac {1}{2}
t_2 (1 + x_2) \rho} \; .
\label{eq:masspola}
\end{equation}
This can be understood if one considers that 
the two body states of fully polarized neutron matter are all 
triplet states on both the spin and the isospin spaces, {\it i.e.,} they are symmetric 
in the spin-isospin variables, and, therefore, due to the Pauli principle, they 
do not see $s$-wave contributions (associated to the purely contact terms 
of the Skyrme force), but only $p$-waves originated from the gradient 
terms. As the parameter $t_2$ is usually taken negative, one has to take 
$ x_2 \leq -1 $ in order to avoid the collapse of fully polarized neutron
 matter \cite{kuts94}. Both SLy4 and SkI3 fulfill this condition. For SLy4, for
instance, $x_2=-1$. For this interaction, totally polarized neutron matter has
$m^*/m=1$ and its energy reduces simply to the kinetic energy of a free Fermi sea.

\begin{figure}[thb]
   \includegraphics[height=10cm,width=12cm]{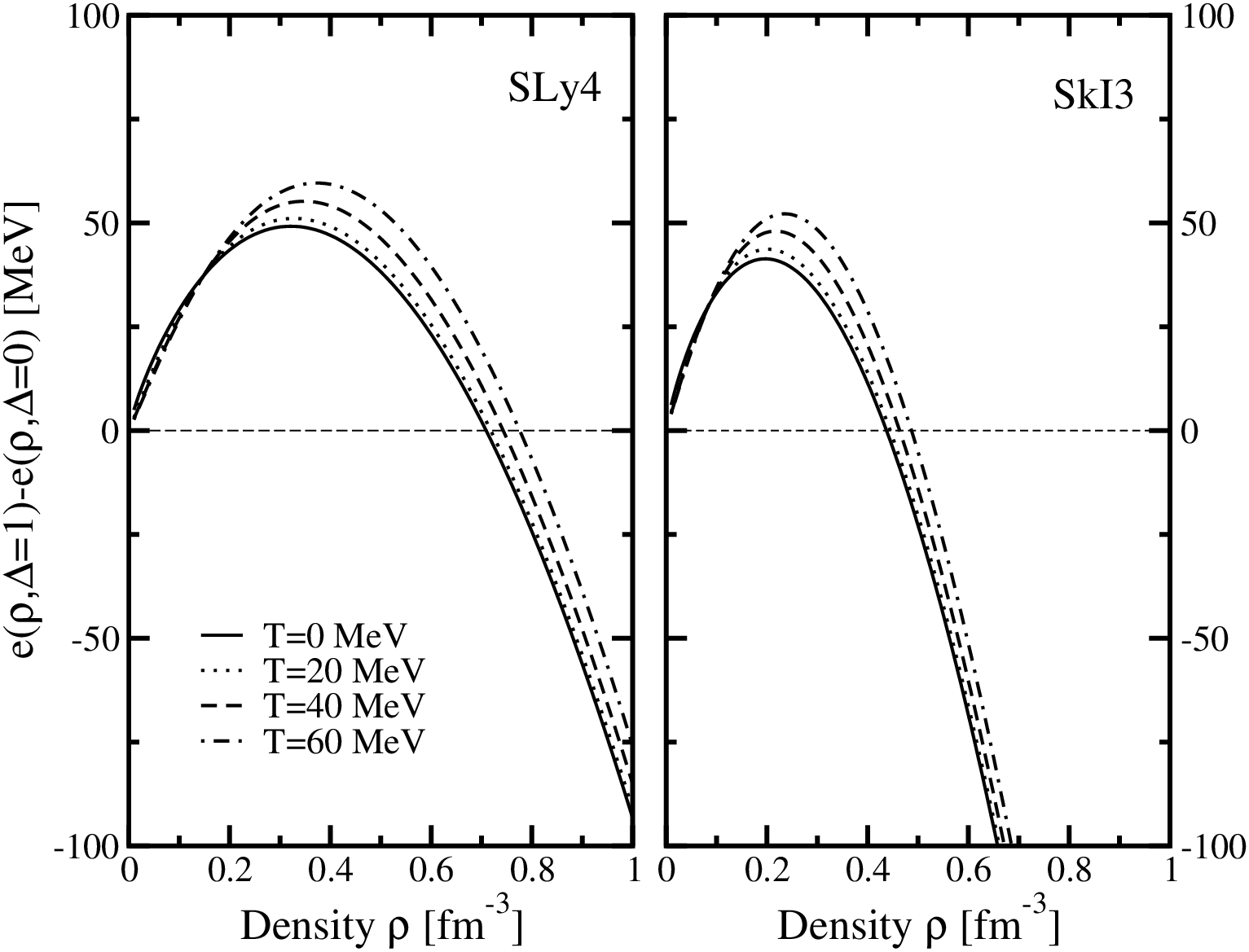}
   \vspace{0.75cm}
   \caption{Difference between the energy per particle of fully polarized neutron matter and unpolarized neutron matter as a function of density for several temperatures.}
   \label{fig:fig3}
\end{figure}

The free energy is composed by the sum of two contributions, the internal energy 
and the entropy. Therefore, in order to understand
the anomalous behaviour of the free energy, it is reasonable to 
analyze how both terms behave as a function of density. With this aim, we show in 
Fig.\ (\ref{fig:fig3}) the internal energy difference $e(\rho,\Delta=1,T)-e(\rho,\Delta=0,T)$ 
as a function of the density for several temperatures. As expected,
the density at which this difference becomes zero grows with temperature, so the origin of
the surprising behaviour detected in Fig.\ (\ref{fig:fig2}) should be attributed to the
entropy contribution of the free energy.  

\begin{figure}[thb]
   \includegraphics[height=10cm,width=12cm]{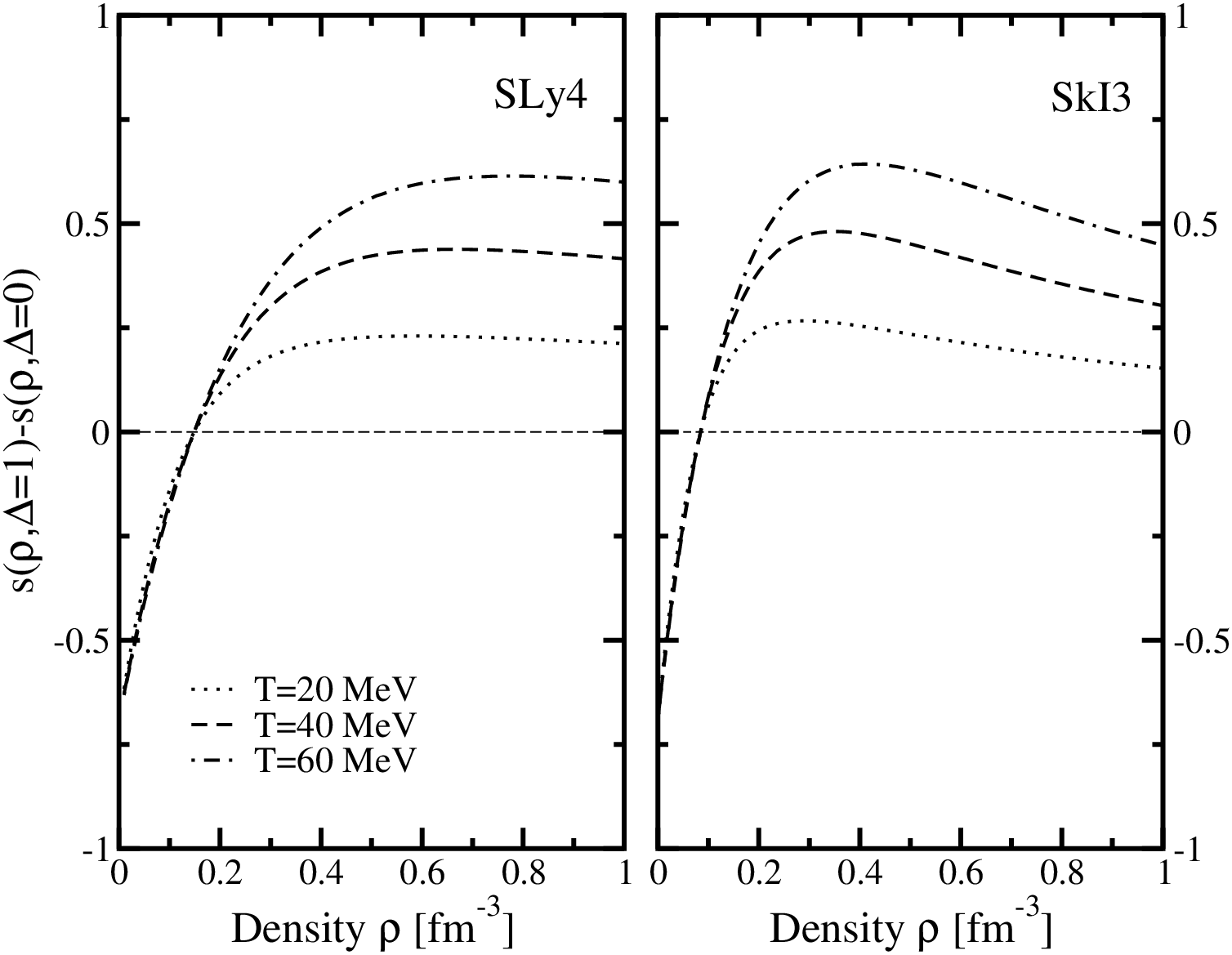}
   \vspace{0.75cm}
   \caption{Difference between the entropy per particle of fully polarized neutron matter and unpolarized neutron matter as a function of density for several temperatures.}
   \label{fig:fig4}
\end{figure}

The difference between the entropy of the polarized and the unpolarized phases
is shown in Fig.\ (\ref{fig:fig4}) as a function of the density for  
three different temperatures.
One would naively expect the entropy of the polarized system to be lower than that of the 
unpolarized one because, intuitively, the fully polarized phase is more ``ordered'' than the
unpolarized one. This is, in fact, the behaviour of the free Fermi sea. However, for the 
interacting system, the modification of the single particle properties, in particular those 
due to the effective mass, can invert this behaviour even in the Hartree-Fock 
approximation. 
This inversion takes place at relatively low densities, 
$\rho_c^S=0.15 \mbox{ fm}^{-3}$ and $\rho_c^S=0.08 \mbox{ fm}^{-3}$ for SLy4 and SkI3 respectively,
much smaller than the onset densities at which ferromagnetism appears. Notice also that the 
entropy critical densities are temperature independent. 

\begin{figure}[thb]
   \includegraphics[height=10cm,width=12cm]{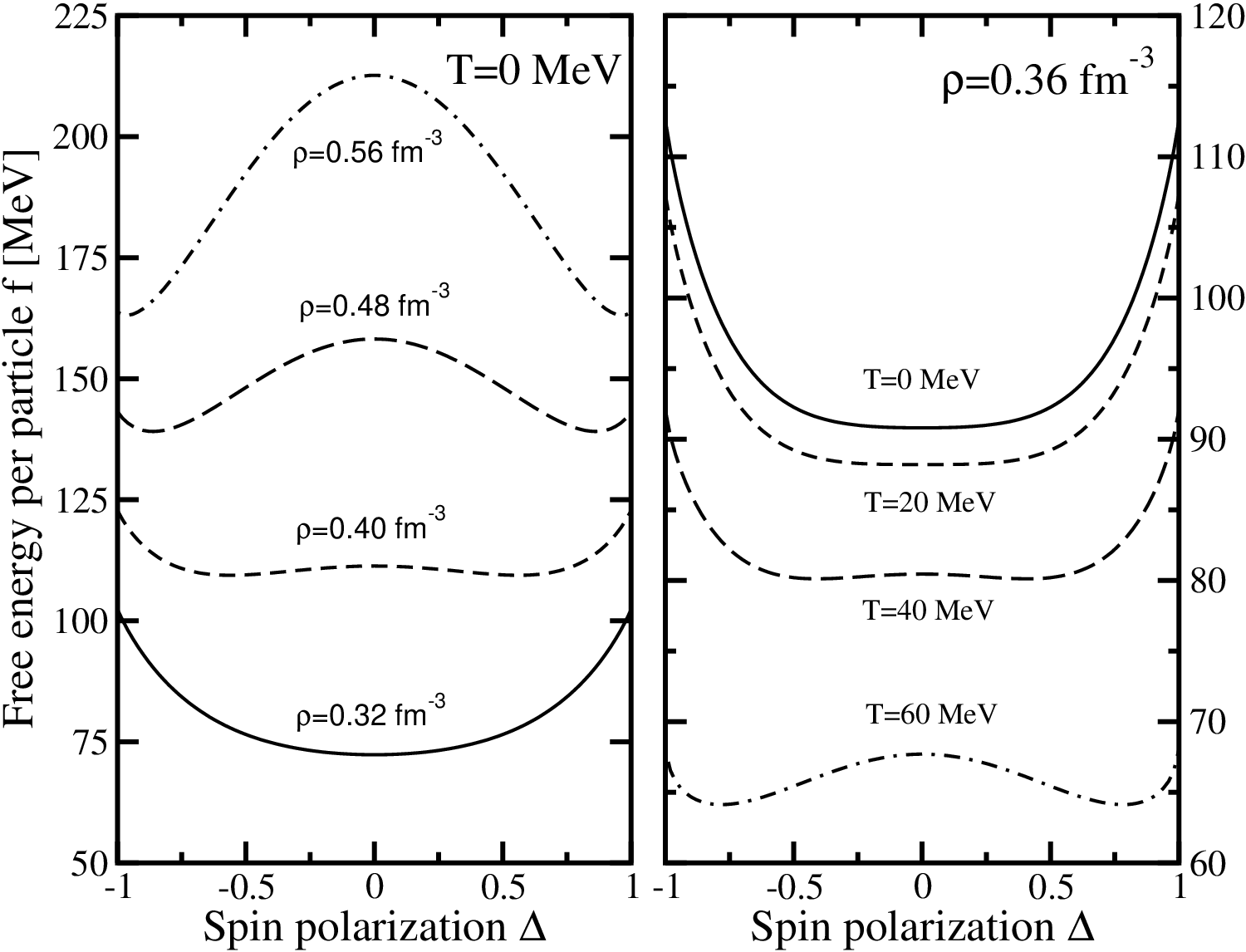}
   \vspace{0.75cm}
   \caption{Left panel: neutron matter free energy per particle at zero temperature as a function
            of polarization for several densities. Right panel: neutron matter free energy per particle at a fixed density $\rho=0.36 \mbox{ fm}^{-3}$
			as a function of polarization for several temperatures. Both figures were obtained using the SkI3 force.}
   \label{fig:fig5}
\end{figure}

As we have seen previously, there is a certain range of densities (between $\rho_c$ and $\rho_c^F$)
where neutron matter is unstable around $\Delta=0$ but the fully polarized phase is still not 
energetically favourable. In this range (and also for densities a bit higher than $\rho_c^F$),
one can see that the minimum free energy happens at a partial polarization $0<\Delta<1$.
This is seen in the left panel of Fig (\ref{fig:fig5}), where we show the free energy 
of the system as a function of polarization at zero temperature (which 
of course coincides with the energy per particle) and for several 
densities, obtained with the SkI3 interaction. At densities below $\rho_c$ (the 
critical density provided by the susceptibility criterium), the minimum energy 
takes place at $\Delta=0$ (this is the case of the first considered density, 
$\rho=0.32 \mbox{ fm}^{-3}$). When we overpass $\rho_c$, the minimum appears at an intermediate 
polarization and the energy of the fully polarized phase is still larger than the 
unpolarized one (this would be the case for $\rho=0.40 \mbox{ fm}^{-3}$).
 When the density increases, the minimum moves to larger values of $\Delta$,
 and, beyond $\rho_c^F$, we observe that $f(\rho,\Delta=0) > f(\rho,\Delta=1)$ (this is the
case for $\rho=0.48 \mbox{ fm}^{-3}$) even though the $\Delta=1$ phase is not the preferred one. Finally, 
for larger densities, the minimum shows up at the fully polarized configuration.  

One can also observe the appearance of a preferred partial polarization as a function
of the temperature for a given density, as it is illustrated in the 
right panel of Fig (\ref{fig:fig5}) where the free energy per particle at $\rho=0.36 \mbox{ fm}^{-3}$ 
is drawn as a function of the polarization. At this density, the internal energy has the minimum 
at $\Delta=0$ for all the temperatures. However, the chosen density is above 
the onset density for the anomalous behaviour of the entropy $\rho > \rho_c^S$, {\it i.e.,} the entropy 
is larger for the polarized phase. As a result, the free energy develops a minimum at $0<\Delta<1$ that
moves to higher polarizations when the temperature increases. Both panels of Fig.\ (\ref{fig:fig5}) are
typical examples of a spontaneous symmetry breaking, where the interaction drives the ground state of
the system to non-zero polarizations. 

\begin{figure}[thb]
   \includegraphics[height=10cm,width=12cm]{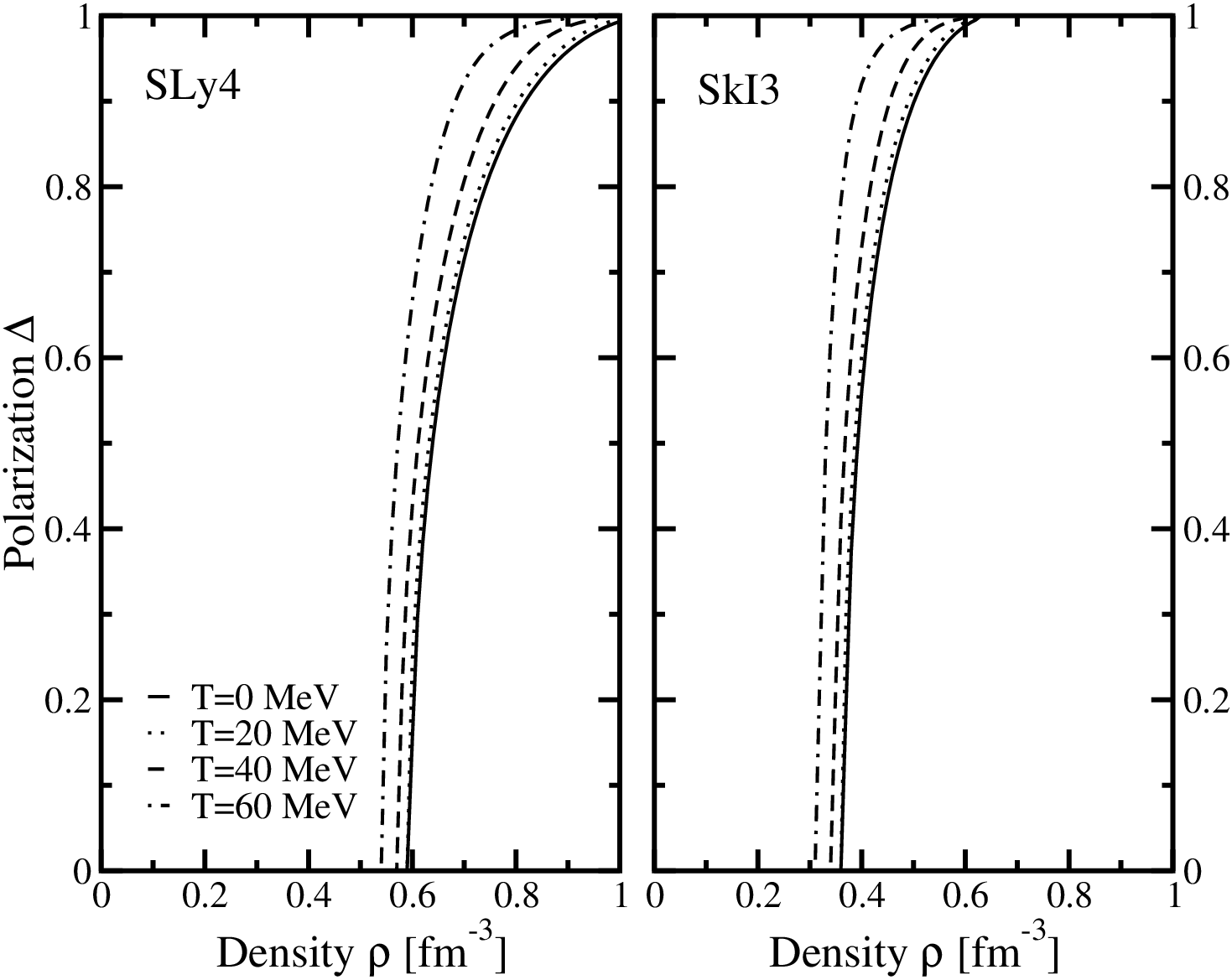}
   \vspace{0.75cm}
   \caption{Polarization associated to the minimum of the free energy for neutron matter  
            as a function of  density for several temperatures. }
   \label{fig:fig6}
\end{figure}

The spontaneous breaking of the spin rotational symmetry indicates the appearance of a phase transition in the system. In our case, this phase transition is between a non-polarized and a polarized phase and the natural order parameter is, thus, the polarization. In Fig.\ (\ref{fig:fig6}) we report the polarization associated to the minimum of the free energy. For a given density and
temperature, this is the polarization that gives the lowest free energy and it
 indicates, then, the polarization at which our system is thermodynamically stable.
With this figure in hand, we can check all the results that we have been discussing up 
to now. At zero temperature, below $\rho_c$, the equilibrium configuration corresponds 
to $\Delta=0$. Over $\rho_c$, the polarization grows steeply up to $\Delta=1$. When the temperature 
increases, the curves shift to smaller densities, so $\rho_c$ decreases with 
temperature and also the density at which the system becomes fully polarized. Of course, 
these results are somewhat academic since the densities needed for neutron
matter to be fully polarized (in particular for SLy4) are very high and, thus, unattainable in
a proto-neutron star. However, we find them useful since they serve to confirm the results concerning
the anomalous behaviour of the free energy with temperature.

\section{Entropy and effective mass}

In this section we investigate the relation between the anomalous behaviour of 
the entropy and the dependence of the effective mass on the polarization. 
To this end, we will determine the density at which the difference between the 
entropy per particle of the fully polarized and the unpolarized phase becomes positive, 
{\it i.e.,} does not show the expected behaviour $s(\rho,T,\Delta=1)-s(\rho,T,\Delta=0) < 0$. We will explore
two situations: the classical limit, defined by the condition 
$ \frac{\rho \lambda^3}{g} \rightarrow 0$ (with $\lambda = \sqrt {\frac{2\pi\hbar^2}{m T } }$ being the 
de Broglie wavelength) and the degenerate limit, where $\frac{T}{\epsilon_F}<<1$. 

In the first place, we consider the classical limit. In the interacting case, it is 
useful to introduce the de Broglie wave length associated to the effective mass:
\begin{equation}
\lambda^* = \sqrt {\frac {2\pi\hbar^2}{m^* T } } \; .
\end{equation}
It is worth reminding that, for Skyrme interactions, the effective mass 
depends on the density but is not affected by the temperature. In addition,
it turns out that one can easily write all the relevant quantities for the 
non-polarized system ($\Delta=0$) and the fully polarized one ($\Delta=1$). For instance, the internal energy can be
casted in the form:
\begin{equation}
e_{cla}(\rho,T,\Delta) = \frac {3}{2} T + U_1(\rho,\Delta) \;, 
\end{equation}
where $U_1(\rho,\Delta)$ is the contribution to the internal energy that 
can not be included into the effective mass terms:
\begin{equation}
U_1(\rho,\Delta)=\frac{\rho}{4} \left[ t_0(1-x_0)+\frac{1}{6}t_3(1-x_3)\rho^{\alpha}  \right] (1-\Delta^2) \; .
\end{equation}
For the entropy, one has:
\begin{equation}
s_{cla}(\rho,T,g) = \frac {5}{2} -\ln 
\left[ \frac {\rho \lambda^{*3}}{g} \right ] \; ,
\end{equation}
where we have explicitly shown the dependence on the degeneracy factor $g$. In order to get
the fully polarized case ($\Delta=1$), it is enough to set $g=1$, while the non-polarized case
($\Delta=0$) is obtained by setting $g=2$. The previous expression can be splitted in two pieces, 
the entropy of a free gas plus a correction term associated to the effective mass:
\begin{equation}
s_{cla}(\rho,T,g) = \frac {5}{2} - \ln \left [ \frac {\rho \lambda}
{g} \right ] + \frac {3}{2} \ln \left [ \frac {m^*}{m} \right ] \; ,
\end{equation}
where one has to take into account that the effective mass depends on both the density $\rho$
and the polarization $\Delta$.
For the free case, one can in fact check that the expected inequality is fulfilled at all
densities and temperatures: 
\begin{equation}
s_{cla}(\rho,T,\Delta=1) - s_{cla}(\rho,T,\Delta=0) = -\ln 2 < 0 \; .
\end{equation}
In the interacting case, however, this difference becomes:
\begin{equation}
\Delta s_{cla}(\rho) \equiv s_{cla}(\rho,T,\Delta=1)-s_{cla}(\rho,T,\Delta=0) = -\ln 2 + \frac {3}{2} \ln \left [ \frac
{m^*(\rho,\Delta=1)}{m^*(\rho,\Delta=0)} \right ] \; ,
\end{equation}
which clearly shows the influence of the interaction, via the effective mass, to the entropy
difference, and which turns out to be independent of the temperature.
If we require the difference to be negative, the following condition 
should be satisfied:
\begin{equation}
\frac {m^*(\rho,\Delta=1)}{m^*(\rho,\Delta=0)} < 2^{2/3} \; . 
\end{equation}
Taking into account the expressions for the effective mass of the 
polarized and unpolarized phases for the Skyrme interactions 
(\ref{eq:effec}), one gets the following condition for the parameters of
the effective force:
\begin{equation}
\frac { 1 + \frac {2 m}{\hbar^2} \frac {\rho}{8} \left [t_1(1-x_1) + 
3 t_2 (1 + x_2) \right ] }{ 1 + \frac {2 m}{\hbar^2} \frac {\rho}{4}
2 t_2 (1 +x_2) } < 2^{2/3} \; .
\label{eq:cond}
\end{equation}
The previous expression is not well defined whenever the denominator
becomes zero. In fact, one can check that eleven of the twenty seven forces
that passed the tests of \cite{stone2003} have a vanishing denominator
at densities below $0.5 \mbox{ fm}^{-3}$. Since this denominator is nothing
but the inverse effective mass for polarized neutron matter, such a 
singularity shows that these parametrizations were not devised to describe
highly spin and isospin asymmetric matter. For the rest of the forces, one can check that 
$m^*(\rho,\Delta=1)/m^*(\rho,\Delta=0)$ is a monotonically 
increasing function of the density, so we can univocally define a critical
density $\rho_c^{cla}$. For SkI3, for instance, one can use the condition 
(\ref{eq:cond}) to obtain the critical density $\rho_c^{cla}=0.08\mbox{ fm}^{-3}$. 
The case for Skyrme Lyon forces is special, since most of them have $x_2=-1$
(the only exceptions are SLy0 and SLy2). In that simpler case, the previous condition 
reduces to:
\begin{equation} 
1 + \frac{2 m}{\hbar^2}\frac{\rho}{8} t_1(1-x_1) < 2^{2/3} \; .
\end{equation}
Then, as $t_1 (1-x_1) > 0$, one can easily obtain a critical density 
($\rho_c^{cla}=0.15\mbox{ fm}^{-3}$ for SLy4) beyond which one can be sure that,
if the temperature is high enough to reach the classical limit, one will observe
an anomalous behaviour of the entropy. In fact, one can check that the temperature
is not an essential factor since we have seen that $\rho_c^S$ (precisely the density
at which $s(\rho,T,\Delta=1)-s(\rho,T,\Delta=0)$ changes sign) is quite independent of temperature and, thus, this
classical entropy critical densities coincide with the $\rho_c^S$ defined before, $\rho_c^{cla}=\rho_c^S$. 
In Table \ref{table2} we give a list of these densities for the sixteen forces which
have no singular effective mass for neutrons in fully polarized matter 
and compare it to the corresponding ferromagnetism onset densities at zero 
temperature $\rho_c$. 

In second place, we investigate the low temperature limit. In this case, 
we can express the different thermodynamic quantities for the polarized
and the unpolarized  phases in terms of the Fermi momenta $k_F = (6 \pi^2 \rho/g)^{1/3}$ 
and the Fermi energies associated to the effective mass 
$\epsilon_F^*(\Delta) = \frac{\hbar^2 k_F^2}{2 m^*(\Delta)}$. 
In the low temperature regime, the internal energy can be written as (this expression
is only valid for $\Delta=0$ or $\Delta=1$):
\begin{equation}
e_{low}(\rho,T,\Delta) = \frac {3}{5} \epsilon_F^*(\Delta) \left [ 1 + \frac {5 \pi^2}{12}
\left ( \frac {T}{\epsilon_F^*(\Delta) }\right ) ^2 \right ] + U_1(\rho,\Delta) \; ,
\end{equation}
where $U_1$ is the same function defined previously for the classical case and where
we recover the well-known $T^2$ dependence of the internal energy. 
The entropy per particle (again only valid for $\Delta=0$ or $\Delta=1$), on the other hand,
 shows a linear dependence on $T$:
\begin{equation}
s_{low}(\rho,T,\Delta) = \frac {\pi^2 }{2 \epsilon_F^*(\Delta)} T 
= \frac {\pi^2}{3 \rho} N(0) T,
\label{eq:entlowT}
\end{equation}
where we have introduced the density of states at the Fermi surface, 
$N(0) = g m^*k_F/ 2 \pi^2 \hbar^2$. 

At this point, we can perform the same analysis that we made for the 
classical limit. Let us start again by the free case and rewrite 
(\ref{eq:entlowT}) to see the explicit dependence on the spin degeneracy:
\begin{equation}
s_{low}(\rho,T,\Delta=1) -s_{low}(\rho,T,\Delta=0) = \frac {\pi^2 T m}{\hbar^2 (6 \pi^2
\rho)^{2/3} } (1 -2^{2/3}) \; ,
\end{equation}
which in fact is smaller than zero. For the interacting case, the only difference 
comes, again, only from the effective masses:
\begin{eqnarray}
\Delta s_{low}(\rho,T) &\equiv& s_{low}(\rho,T,\Delta=1) - s_{low}(\rho,T,\Delta=0) = \nonumber \\
&=&\frac {\pi^2 T m^*(\rho,\Delta=0)}{\hbar^2 (6 \pi^2 \rho)^{2/3}}
\left[\frac{m^*(\rho,\Delta=1)}{m^*(\rho,\Delta=0)} - 2^{2/3} \right] \; .
\end{eqnarray} 
By requiring this difference to be negative, we recover the same condition for
the effective masses that we found in the classical regime (\ref{eq:cond}). Contrary to
the classical limit, in this case $\Delta s_{low}$ depends on the temperature, although this 
dependence can not change its sign. Now we can understand why $\rho^S_c$ is independent
of the temperature: since in the classical limit (which requires high temperatures)
and in the low temperature limit the entropy critical densities are exactly the same, one
can guess that for intermediate temperature $\rho^S_c$ will also not change.

\begin{figure}[thb]
   \includegraphics[height=10cm,width=12cm]{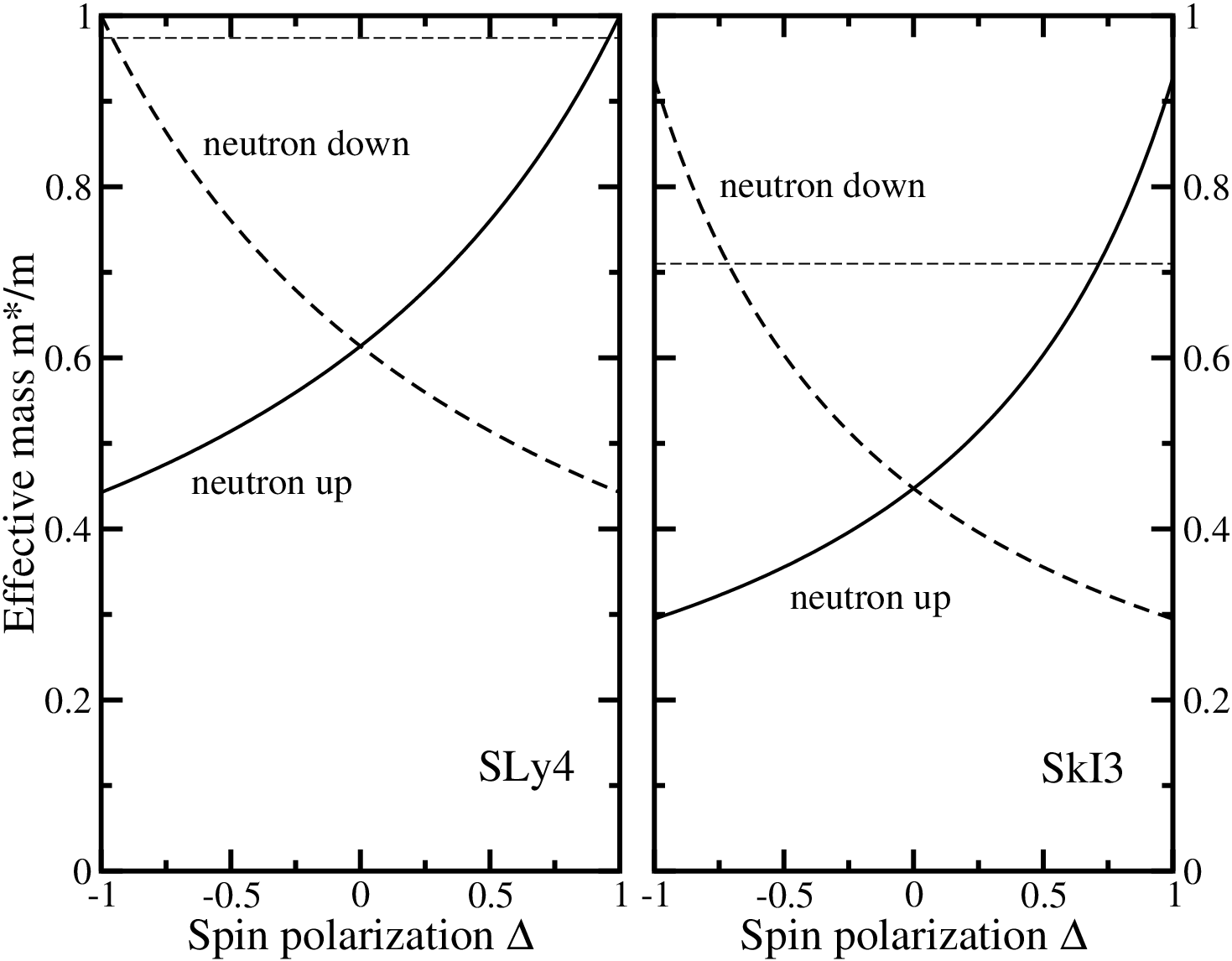}
   \vspace{0.75cm}
   \caption{Neutron effective mass of the up and down components as a function 
            of polarization at  $\rho=0.16 \mbox{ fm}^{-3}$. The horizontal lines signal the 
            maximum value of the effective mass of the up component in fully polarized matter
			if the entropy of the polarized phase has to be smaller than the unpolarized one.}
   \label{fig:fig7}
\end{figure}

We have seen that the interaction influences the entropy only through
the effective mass and we have been able to trace back the anomalous behaviour of the entropy 
(as a function of the polarization) to the violation of relation (\ref{eq:cond}), which does only involve effective
masses. In Fig.\ (\ref{fig:fig7}), then, we show the effective mass of neutrons with
 both orientations of the third spin component as a function of polarization for a fixed density
$\rho= 0.16 \mbox{ fm}^{-3}$, which lies above the entropy critical density $\rho_c^S$ for both SLy4 and SkI3. 
Obviously, for $\Delta=0$, the effective masses for the neutrons with spin up and spin down coincide.
The horizontal line in both panels is placed at $m^*/m = 2^{2/3}m^*(\Delta=0)/m $,
indicating the upperbound for $m^*(\Delta=1)/m$, above which we will find the anomalous behaviour 
of the entropy. The effective mass of the most abundant species increases with the polarization,  
while the one of the less abundant is the decreasing one. Notice that for SLy4, as we have already
pointed out, the effective mass for the fully polarized system is simply the bare mass. 

\begin{figure}[thb]
   \includegraphics[height=10cm,width=12cm]{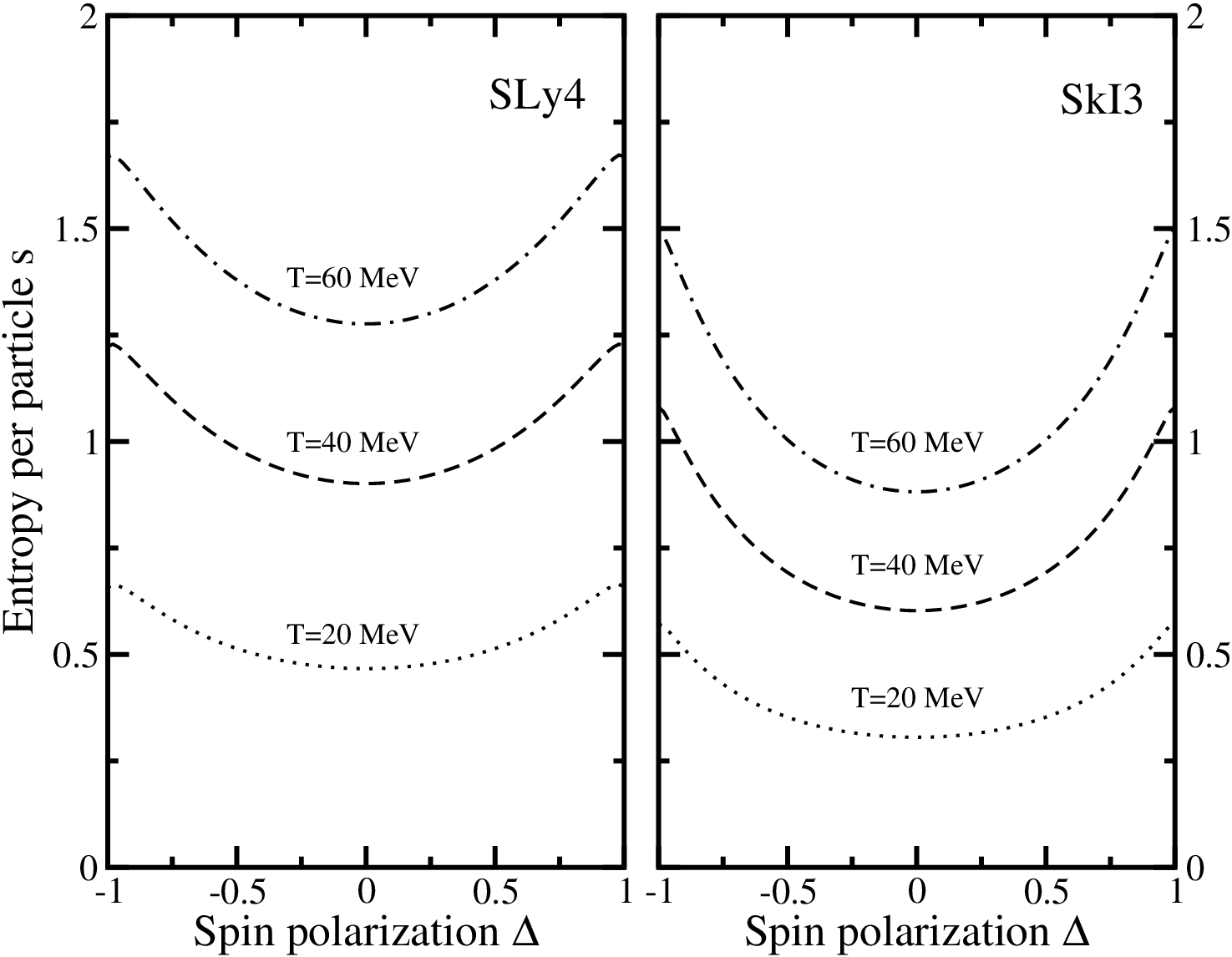}
   \vspace{0.75cm}
   \caption{Entropy per particle of neutron matter as a function of polarization at 
            $\rho=0.32 \mbox{ fm}^{-3}$ for several temperatures. }
   \label{fig:fig8}
\end{figure}

The behaviour of the entropy as a function of the polarization at a fixed density $\rho=0.32 \mbox{ fm}^{-3}$ 
for several temperatures and for both Skyrme forces is shown in Fig.\ (\ref{fig:fig8}). As 
expected, the entropy increases with temperature. However, since 
this density is above $\rho_c^S$, the entropy of the fully polarized phase is larger
than that of the unpolarized one, giving the anomalous behaviour of the entropy with the polarization.
 A careful look shows that this difference increases 
with temperature as expected from the results of Fig.\ (\ref{fig:fig4}). For a given temperature and 
polarization and considering that the entropy in both limits
(and also in between them) is an increasing function of the effective 
mass and that, as we have seen in Fig.\ (\ref{fig:fig7}), the effective mass for SLy4 is larger than 
that of SkI3, we can understand why the entropy associated to SLy4 (left panel) is larger than the one
corresponding to SkI3 (right panel).

Finally, using the low temperature expansion, we will analyze why 
$\rho_c^F$ decreases with temperature. 
At zero temperature, the difference between the free energy (internal energy) of the fully polarized phase
and the non-polarized one is:
\begin{eqnarray}
\Delta f(\rho,T=0) & \equiv & f(\rho,T=0,\Delta=1) - f(\rho,T=0,\Delta=0) = \nonumber \\
&=& \frac {3}{5} \frac {\hbar^2}{2 m^*(\rho,\Delta=1)}
(3 \pi^2 \rho)^{2/3} \left [ 2^{2/3} - \frac {m^*(\rho,\Delta=1)}{m^*(\rho,\Delta=0)}
\right ] \nonumber \\ 
&&- \frac {1}{4} \rho \left [ t_0 (1-x_0) + \frac {1}{6} t_3 (1-x_3) \rho^{\alpha}
\right ] \; .
\end{eqnarray}
In order to find $\rho_c^F(T=0)$, we should find the density at which the previous expression 
becomes zero. 
For most of the forces, as we have already shown in Table \ref{table2}, the density at which 
the first term becomes negative is quite low, while the combination 
$\left [ t_0 (1-x_0) + \frac {1}{6} t_3 (1 -x_3) \rho^{\alpha} \right ]$ 
is usually negative at all densities. 
The energy difference, then,  
becomes negative only through a balance of both terms, and 
this happens at $\rho=\rho_c^F(T=0)$, which is an 
upperbound for the real critical density $\rho_c(T=0)$. 
It is quite noticeable
 that this cancellation results from a subtle interplay of all parameters of the Skyrme force.
Once we introduce a low temperature, the difference between free energies becomes:
\begin{eqnarray}
f(\rho,T,\Delta=1)- f(\rho,T,\Delta=0) &=& \Delta f(\rho,T=0) \nonumber \\
&& - \frac {\pi^2}{4} T^2 \frac {2 m^*(\rho,\Delta=0)}{\hbar^2 (6 \pi^2 \rho)^{2/3}}
\left [ \frac {m^*(\rho,\Delta=1)}{m^*(\rho,\Delta=0)} - 2^{2/3} \right ]
\end{eqnarray}
and, therefore, this relation shows that at finite temperature,
whenever $m^*(\rho,\Delta=1)/m^*(\rho,\Delta=0) > 2^{2/3}$, which occurs 
for densities larger than $\rho_c^S$, the unpolarized system is unstable 
if the instability happens already at $T=0$. However, even when 
$\Delta f(\rho,T=0)>0$, the second term can make the difference become 
negative. Around $\rho_c^F(T=0)$, the difference decreases with density 
(see Fig.\ (\ref{fig:fig2})) and thus the negative term makes $\rho_c^F(T \neq 0)$ 
smaller with respect to $\rho_c^F(T=0)$. Or, in other words, 
$\rho_c^F(T) < \rho_c^F(T=0)$ provided that 
$\rho_c^F(T=0) > \rho_c^S$, which is always true as seen in Table \ref{table2}.

\section{Summary and conclusions}

In this work, we have studied the properties of polarized neutron matter, 
with neutrons interacting through Skyrme-type interactions, both at zero and
finite temperature. Firstly, we have revised the zero temperature
calculations using two modern Skyrme forces and we have studied the ground state
of neutron matter as a function of polarization. We have shown that the 
ground state is not necessarily at fully polarized or unpolarized matter,
but that it can be found at partially polarized matter, giving rise to a 
spontaneous symmetry breaking where the system prefers a state with non-zero
polarization. 

Our main emphasis, however, has been the study of the influence of temperature 
on the manifestation of the ferromagnetic behaviour. In particular, we have considered
the stability of the unpolarized phase and we have shown that the critical density at
which ferromagnetism takes place $\rho_c$ decreases with temperature. This unexpected
behaviour has been associated to an anomaly of the entropy:
above a certain critical density $\rho_c^S$, the entropy of the polarized phase turns 
out to be larger than that of the unpolarized one. We have also shown that this fact
is a consequence of the dependence of the entropy on the effective mass of the 
neutrons with different third spin component and, in particular, a consequence of the
dependence of these effective masses on the polarization.
More precisely, we have derived a condition for the maximum ratio between the 
effective masses of the fully polarized phase and the unpolarized one. Although this
criterium is density dependent, one could in principle use it as a restriction on the
parameters defining a Skyrme force. 

Finally, we would like to emphasize the fact that the present analysis has been 
restricted to Skyrme interactions which, contrary to other microscopic calculations, 
give a ferromagnetic transition at densities around $3.5\rho_0$. It would be useful to
compare these calculations with the results obtained with realistic interactions
and determine the behaviour of the entropy and the effective masses of neutrons as 
a function of the spin polarization. Work in this direction is presently in progress.

\vspace{0.5cm}

{\bf Acknowledgments}

The authors are very grateful to Profs. A. Ramos, J. Navarro, I. Bombaci and J. Ort\'\i n for 
useful and stimulating discussions. This research was also partially supported by DGICYT (Spain) 
Project No.\ BFM2002-01868 and from Generalitat de Catalunya Project 
No.\ 2001SGR00064. One of the authors (A. R.) acknowledges the support from DURSI and 
the European Social Funds.

\newpage

\begin{table}
\begin{tabular}{|c|c|c|c|c|c|} \hline
 & $\rho_{\infty}$ (fm$^{-3}$) & $a_v$ (MeV) & $a_s$ (MeV) & $K_{\infty}$ (MeV) & $M_{max}$ ($M_{\odot}$)   \\ \hline
SLy4 & 0.160 & -15.97 & 32.04 & 230.9 & 2.04 \\ \hline
SkI3 & 0.158 & -15.98 & 34.89 & 259.2 & 2.19 \\ \hline
\end{tabular}
\vspace{1cm}
\caption{Symmetric Nuclear Matter properties at saturation density for SLy4 and SkI3. 
The corresponding value of the maximum mass of a neutron star obtained with the EoS of these two forces is also shown.} 

\label{table1}
\end{table}

\newpage

\begin{table}
\begin{tabular}{|c|c|c|} \hline
		& $\rho_c^S$ (fm$^{-3}$)	& $\rho_c$ (fm$^{-3}$) \\  \hline
SGI 	& 0.09						& 0.28		\\ \hline
SLy0 	& 0.11						& 0.41		\\ \hline
SLy1 	& 0.15						& 0.58		\\ \hline
SLy10 	& 0.13						& 0.61		\\ \hline
SLy2 	& 0.09						& 0.29		\\ \hline
SLy230a	& 0.10						& 0.54		\\ \hline
SLy3 	& 0.15						& 0.60		\\ \hline
{\bf SLy4} & 0.14					& 0.60		\\ \hline
SLy5 	& 0.15						& 0.57		\\ \hline
SLy6 	& 0.14						& 0.90		\\ \hline
SLy7 	& 0.14						& 0.57		\\ \hline
SLy8 	& 0.15						& 0.60		\\ \hline
SLy9 	& 0.11						& 0.47		\\ \hline
SV	 	& 0.15						& 0.77		\\ \hline
{\bf SkI3} & 0.08					& 0.37		\\ \hline
SkI5 	& 0.07						& 0.28		\\ \hline
\end{tabular}
\vspace{1cm}
\caption{Densities $\rho_c^S$ at which the condition (\ref{eq:cond}) is violated for 16 different Skyrme forces. The respective critical densities $\rho_c$ for the onset of ferromagnetism at T=0 are also shown.}
\label{table2}
\end{table}

\newpage

\newpage

\newpage

\newpage

\newpage

\newpage

\newpage

\newpage

\end{document}